\RequirePackage{fix-cm}
\documentclass[epjc3]{svjour3}
\smartqed  
\RequirePackage{graphicx}
\RequirePackage{subfigure}
\RequirePackage{amsmath}
\RequirePackage{amssymb}
\RequirePackage{amsfonts}

\RequirePackage[numbers,sort&compress]{natbib}
\RequirePackage[colorlinks,citecolor=blue,urlcolor=blue,linkcolor=blue]{hyperref}
\RequirePackage{xcolor}

\journalname{Eur. Phys. J. C}
\graphicspath{{figure/}}

\usepackage{lineno}

\definecolor{darkgreen}{rgb}{0.0, 0.5, 0.0}

\begin{document}

\title{Impact of the PDFs on the $Z$ and $W$ lineshapes at LHC
}

\author{Valerio Bertacchi\thanksref{vberta:infn,add:SNS, add:INFN}
        \and
        Lorenzo Bianchini\thanksref{add:INFN}
        \and
        Elisabetta Manca\thanksref{add:SNS, add:INFN}
        \and
        Gigi Rolandi\thanksref{add:SNS,add:INFN}
        \and
        Suvankar Roy Chowdhury\thanksref{add:SNS,add:INFN}
}

\thankstext{vberta:infn}{e-mail: valerio.bertacchi@pi.infn.it}

\institute{Scuola Normale Superiore di Pisa, Italy \label{add:SNS}
           \and
           INFN sezione di Pisa, Italy \label{add:INFN}
}

\date{Received: date / Accepted: date}

\maketitle

\begin{abstract}
The parton distribution functions (PDFs) of the proton play a role in determining the lineshape of  $Z$ and $W$ bosons produced at the LHC. In particular, the mode of the gauge boson virtuality is shifted with respect to the  pole due to the dependence of the partonic luminosity on the boson virtuality. The knowledge of this shift contributes to the systematic uncertainty for a direct measurement of the boson mass. A detailed study of the shift and of its systematic uncertainty due to the limited knowledge of the PDFs is obtained using a tree-level model of  $Z$ and $W$ boson production in proton-proton collisions at $\sqrt{s}=13$~TeV.  A Monte Carlo simulation is further used to validate the tree-level model and study the dependence of the shift on the transverse momentum of the  gauge bosons. The tree-level calculation is found to provide a good description of the shift. The systematic uncertainty on the lineshape due to the PDFs is estimated to be below one MeV in the phase-space relevant for a future high-precision mass measurement of the gauge boson masses at the LHC.
\end{abstract}

\section{Introduction}\label{sec:intro}

The masses of gauge bosons are some of the most relevant observables to test the electroweak theory and they are measured with high precision. The $Z$ mass was extracted at LEP from a fit to the lineshape, i.e. from the cross section of the process $e^+e^-\to\gamma,Z\to f\overline f$ measured with different beam energies  spanning center-of-mass energies near $\sqrt{s}\sim M_Z$.  At LEP, the lineshape was distorted by the initial state radiation of the colliding electrons, which was theoretically very well understood, resulting in a systematic uncertainty of less than 0.1 MeV on the $Z$ mass. The  $Z$ mass was ultimately measured with an uncertainty of 2.1 MeV \cite{LEP:MZ} where the largest contribution was the precision of the beam energy calibration.

The experimental situation at CERN's Large Hadron Collider (LHC) differs by at least two aspects. First, proton-proton collisions at a center-of-mass-energy $\sqrt{s}$ result in partonic $f\overline f$ collisions with a broad distribution of partonic center-of-mass-energies  $\sqrt{\hat{s}}=\sqrt{x_1x_2s}$, where $x_1$ and $x_2$ are the fractions of the proton momentum carried by the interacting partons, as described in the empirical \textit{parton density functions} (PDFs). The center of mass of the initial state is unknown and changes on an event by event basis covering all relevant energies. The lineshape of the boson can be measured using the decay product kinematics, like the invariant mass of the $\mu^+\mu^-$ pairs in $Z$ decays.

Secondly, at the LHC the gauge bosons are produced at a rate that is several orders of magnitude larger than LEP one.  The ATLAS and CMS experiment already collected about of 400 millions $W$ and 40 millions $Z$ bosons each during the Run 2 of LHC. By the end of the Run 3,  $10^8$  $Z$ leptonic decays will be available and a factor of 10 more at the end of the High Luminosity LHC program. These numbers must be compared to the 10 millions $Z$ produced at LEP in all decay channels. This unprecedented number of $W$ and $Z$ bosons produced at the LHC offers new opportunities for precise measurements, but it also forces consideration of sources of systematic uncertainty which may have been neglected so far, for example those related to the modeling of the virtuality of the gauge bosons~\cite{Stirling:PDF}. For instance, by analyzing the dileptonic $Z$ decays of Run 3 of the LHC a statistical-only precision of about $10^{-5}\;\text{GeV}$ on $M_Z$ might be achievable. This level of precision would demand a control of the dilepton mass lineshape at the sub-MeV level.

A thorough assessment of the systematic uncertainty on the lineshape is important because the $Z$ mass is used to calibrate the muon momentum scale of the detectors \cite{MomentumScale,Atlas:Wm}. In addition this information may be useful for a new, more precise,  measurement of the $Z$ mass in case the momentum scale of the detector could be calibrated to a relative uncertainty of $10^{-5}$ using independent experimental information, like the $J/\psi$ mass which is known with a relative uncertainity of $10^{-6}$~\cite{PDG}, and assuming that final state radiation effects can be understood to this level of precision.

At first order, the distribution of the virtuality $Q$ of a gauge boson ($V$) originates from the convolution of a relativistic Breit-Wigner with the partonic luminosity function, see e.g. Ref.~\cite{Stirling:QCDpink}. The latter is a function of the dimensionless parameter $\tau = Q^2/s$. The non-trivial dependence of the partonic luminosity on $\tau$ implies a distortion of the lineshape compared to a pure Breit-Wigner. Given the narrowness of the electroweak gauge boson width $\Gamma_V$, this effect can be treated, in first approximation, as a shift $\Delta_V$ of the mode of the distribution compared to $M_V$. The limited knowledge of the PDFs introduces an uncertainty on $\Delta_V$, which contributes directly to the model uncertainty in the extraction of $M_V$ from the dilepton mass distribution.

The goal of this paper is to assess the size of this shift and of its uncertainty due to the limited knowledge of the PDFs. This shift can be regarded as a proxy of the systematic uncertainty on the $Z$ mass extracted from the fit to the dilepton mass distribution.

The $W$ boson lineshape is also distorted by the same effect. However, this is mostly of academic interest since the invariant mass of the leptonic final state cannot be measured in $W$ decays due to the presence of the neutrino. The traditional measurement of the $W$ mass at hadron colliders uses non-Lorentz invariant quantities (e.g. the transverse mass or the lepton transverse momentum) whose distributions have a dependence on the PDFs which induces a systematic uncertainties much larger than the effect discussed in this paper \cite{Vicini:PDFUnc}.
However, the larger than ever amount of $W$ decays collected by the LHC opens up possibilities for novel measurements that might be less sensitive to the PDFs \cite{Bianchini:agnostic,Vicini:Unc}. Likewise, the increased coverage in rapidity planned by the upgrades of the LHC experiments may have a strong impact in reducing the aforementioned PDF uncertainty \cite{Vicini:forwardW, HLLHC:PDF}. For the above reasons, the results of this paper may become relevant also for a future $W$ mass measurement at LHC. 

This paper is organized as follows. In Sec.~\ref{sec:lhapdf} a tree-level calculation of the shift $\Delta_V$ is presented. In Sec.~\ref{sec:MC}, a Monte Carlo (MC) simulation of Drell-Yan production is used to validate the tree-level model and extend the study to the full phase-space. The results are summarized in Sec.~\ref{sec:conclusion}.

\section{Tree level study}\label{sec:lhapdf}

\begin{figure}[!bt]
\centering
\subfigure[\textit{}\label{fig:diagram_tree}]{\includegraphics[width=0.3\columnwidth]{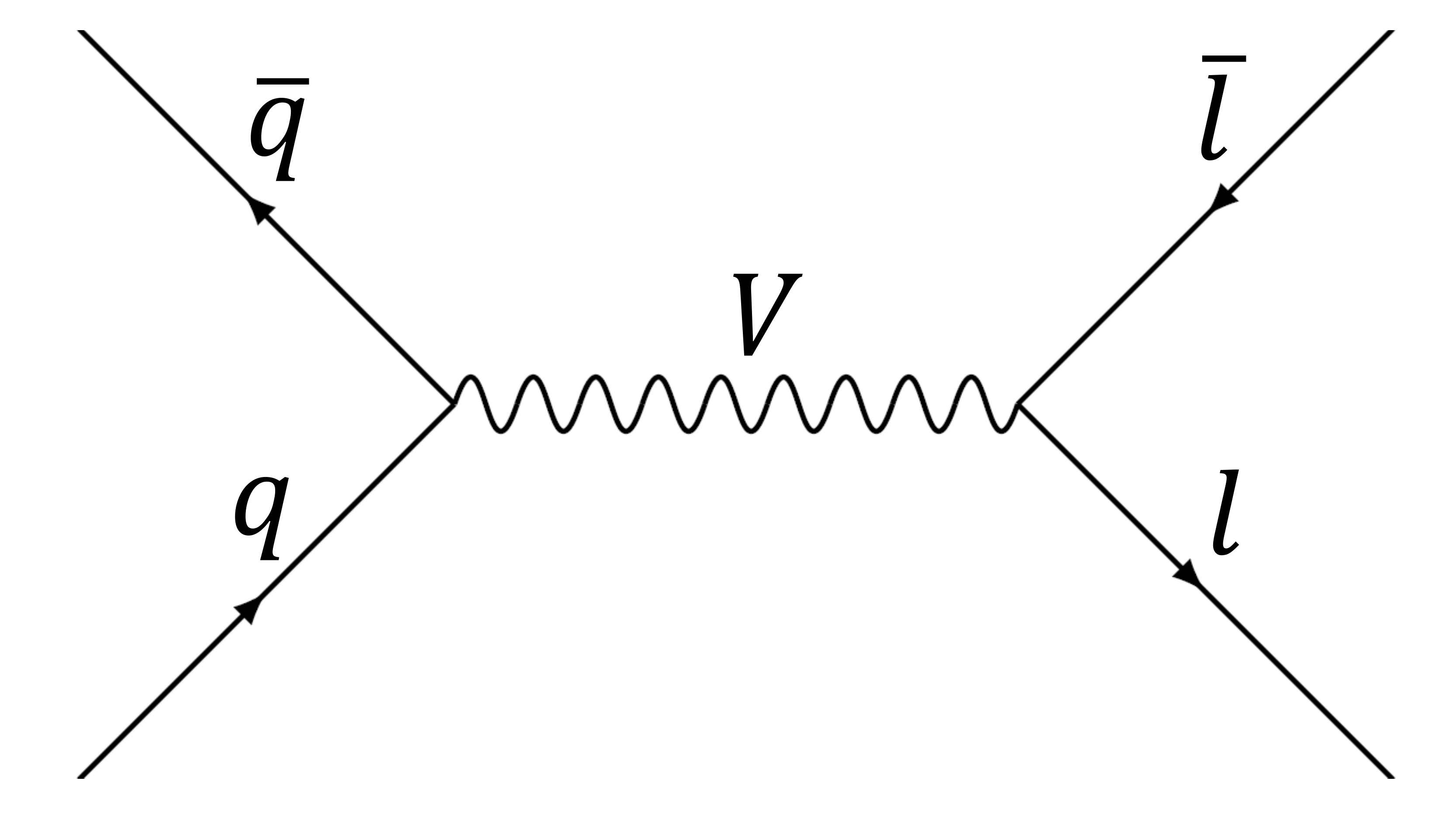}}
\hspace{ 0.2 cm}
\subfigure[\textit{}\label{fig:diagram_qqg}] {\includegraphics[width=0.3\columnwidth]{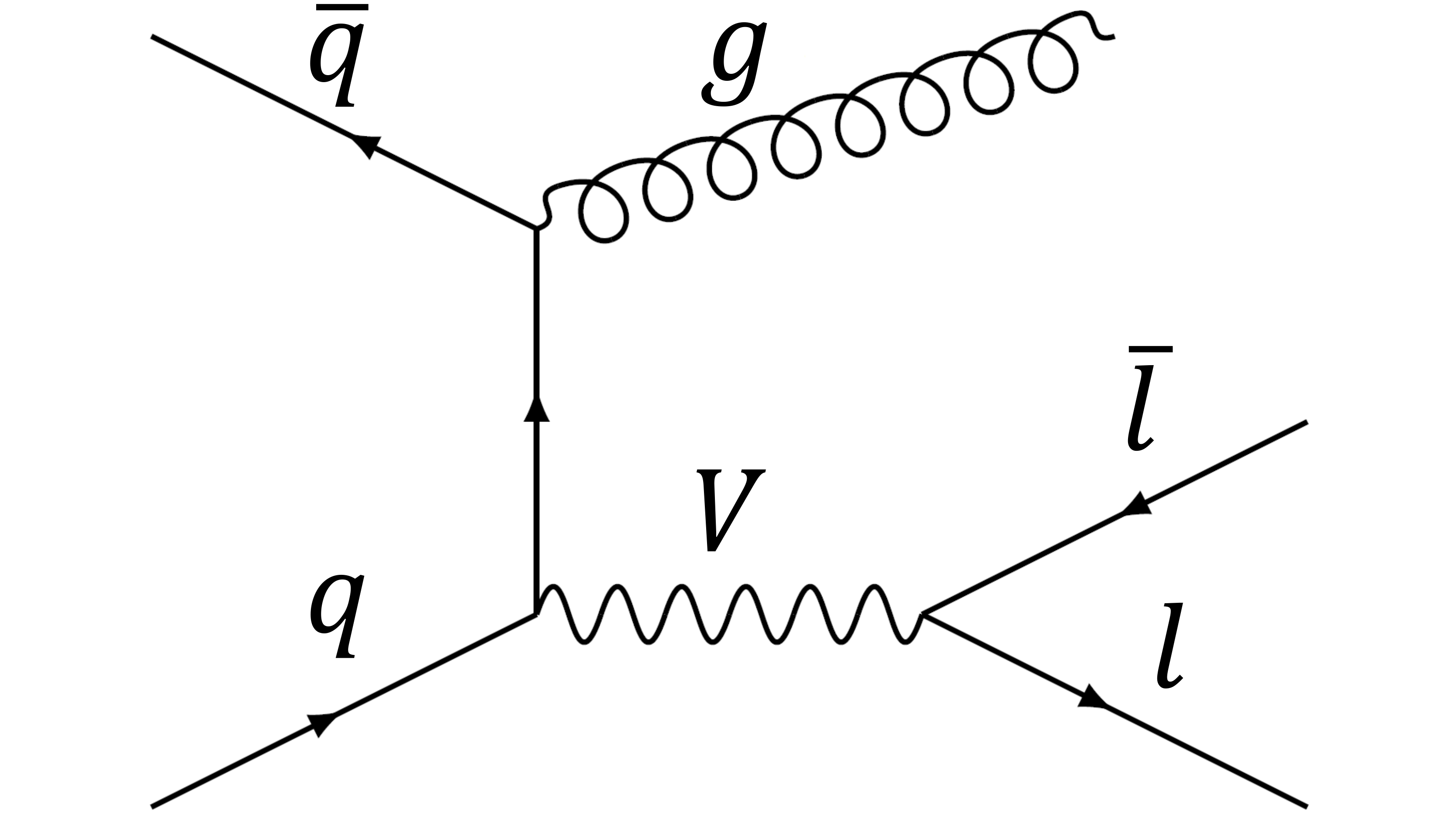}}
\caption{Tree level diagram (\ref{fig:diagram_tree}) and example of NLO diagram (\ref{fig:diagram_qqg}) for vector boson $V$ production.}\label{fig:diagrams}
\end{figure}

A simplified model of Drell-Yan production is first considered based on a minimal set of tree-level diagrams. This approximation amounts to considering just one Feynman diagram per quark-antiquark pair, as illustrated in Fig.~\ref{fig:diagram_tree}. Besides accounting already for the bulk of the total cross section (about $80\%$ for a $20$~GeV threshold on the transverse momentum of the extra parton at $\sqrt{s}=13$ TeV), these diagrams are also expected to be the most sensitive to the PDF-dependent shift under study. Indeed, they are the only $2\to1$ diagrams contributing to the amplitude, whereas higher-order diagrams are at least $2\to2$, see e.g. Fig.~\ref{fig:diagram_qqg}. As such, they include additional invariants besides $Q$.  The existence of these extra scales is expected to dilute the sensitivity of the lineshape to the details of the PDFs. This assumption will be validated by a MC analysis of $pp\to V+X$ production discussed in Sec.~\ref{sec:MC}. In the following,  the \texttt{NNPDF3.0}~\cite{NNPDF} PDF set will be used to evaluate the PDFs relevant for $W$ and $Z$ production in proton-proton collisions at $\sqrt{s}=13$~TeV.

Within the tree-level approximation, the double-differential cross section for $pp\to V(\to \ell\ell^\prime)+X$, as a function of the quark momentum fractions $x_{1,2}$, is given by
\begin{equation}\label{eq:start_xsec}
\frac{d^2\sigma_V}{dx_1dx_2} = \frac{1}{ N_{\rm C}}\sum_{ij}  \bigl[f_i(x_1)f_j(x_2)+f_i(x_2)f_j(x_1)\bigl] \frac{16 \pi \Gamma_V^2  \mbox{BR}_{V\to q_i\overline q_j}\mathrm{B}_{V\to \ell \ell^\prime}}{(x_1x_2s-M_V^2)^2+M_V^2\Gamma_V^2},
\end{equation}
where $N_{\rm C}$ is the number of QCD colours, $M_V$ and $\Gamma_V$ are the mass and width of the resonance, $\mathrm{BR}_{V\to ab}$ are the relevant branching fractions, and the sum at the right-hand side runs over the different combinations of quark flavours contributing to the process under study. For simplicity, the scale-dependence is omitted from the quark PDF $f_{i}(x)$.
In Eq.~\ref{eq:start_xsec} a relativistic Breit-Wigner function with fixed width has been assumed, which is also the functional form used for the MC simulation discussed in Sec.~\ref{sec:MC}. The opportunity of using a running width scheme~\cite{PDG} has been studied as well. However, since the interesting feature in Eq.~\ref{eq:start_xsec} concerns the core of the distribution, where $Q\sim M_V$, the results would not change because the two schemes differs only when $|Q-M_V|\gg \Gamma_V$.

In order to express the double-differential distribution of Eq.~\ref{eq:start_xsec} as a function of $Q$ and $y$ the canonical transformation
\begin{equation}\label{eq:change_var}
y= \frac{1}{2}\ln \frac{x_1}{x_2}, \;\;\;\ Q^2= x_1x_2s,
\end{equation}
 is performed. By combining Eq.~\ref{eq:start_xsec} and \ref{eq:change_var}, the single-differential distribution $d\sigma_V/dQ$, conditional on the rapidity $y$, is obtained:
\begin{equation}\label{eq:full_formula}
\small{
\begin{split}
 \frac{d\sigma_V}{dQ}(Q \; | \; y) & = \left(\frac{d\sigma_V}{dy}\right)^{-1}\frac{1}{N_C}\sum_{ij}\frac{8Q}{s} \bigl[f_i(\bar x_1)f_j(\bar x_2)+f_i(\bar x_2)f_j( \bar x_1)\bigl] \frac{16 \pi \Gamma_V^2  \mbox{BR}_{V\to q_i\overline q_j}\mathrm{B}_{V\to \ell \ell^\prime} }{(Q^2-M_V^2)^2+M_V^2\Gamma_V^2}
 \\
 &\equiv \sum_{ ij}  C^V_{ij}  \bigl[f_i( \bar x_1)f_j(\bar x_2)+f_i(\bar x_2)f_j(\bar x_1)\bigl] \frac{Q}{(Q^2-M_V^2)^2+M_V^2\Gamma_V^2},
 \end{split}
 }
\end{equation}
where $\overline x_{1,2}=\sqrt{\tau}e^{\pm y}$ and the constants $C^V_{ij}$ include terms that depend on $y$ but not on $Q$.
The fact that $\Gamma_V/M_V\ll 1$ and that $f_{i}$ are smooth functions in the relevant range of Bjorken $x$ values ($10^{-3}\lesssim x \lesssim 10^{-1}$) can be exploited to perform a Taylor expansion of the right-hand side of Eq.~\ref{eq:full_formula} aroud $Q=M_V$:
{\small{
\begin{align}
\label{eq:second_order_formula}
& \frac{d\sigma_V}{dQ}(Q \; | \; y) \sim \frac{Q}{(Q^2-M_V^2)^2+M_V^2\Gamma_V^2} \sum_{ij} |V_{ij}|^2\left( F^{ij}+ F^{ji}\right) \times \\ \nonumber
& \left[1 + \underbrace{\frac{\sum_{ ij} |V_{ij}|^2\left( F^{ij}H^{ij}+ F^{ji}H^{ji} \right) }{\sum_{ ij}|V_{ij}|^2\left( F^{ij}+ F^{ji} \right)}}_{H_V}\left(\frac{Q}{M_V} - 1\right)+  \underbrace{\frac{\sum_{ij}|V_{ij}|^2 \left( F^{ij}K^{ij}+ F^{ji}K^{ji} \right) }{\sum_{ij}|V_{ij}|^2 \left( F^{ij}+ F^{ji} \right) }}_{K_V}\left(\frac{Q}{M_V}-1\right)^2\right]
\end{align}}}
where the flavour-dependent terms have been factored out of $\mbox{BR}_{V\to q_i\overline q_j}$ in the form of the square of the $V$ matrix elements. The latter should be interpreted as the usual CKM matrix for the case of $W$ production, and as $\left(T^3_i-2Q_i\sin^2\theta_W\right)\delta_{ij}$ for $Z$ production, where $T^3_i$ and $Q_i$ are the weak isospin and electric charge of quark $i$, respectively. In Eq.~\ref{eq:second_order_formula}, the following auxiliary functions of $y$ have been introduced:
\begin{align}\label{eq:aux}
F^{ij}  &=  \left[f_i(\bar x_1)f_j(\bar x_2)\right]_{Q=M_V} \\ \nonumber
H^{ij}  &= \left[\frac{f_i^\prime(\bar x_1)}{f_i(\bar x_1)}\bar x_1+\frac{f_j^\prime(\bar x_2)}{f_j(\bar x_2)}\bar x_2\right]_{Q=M_V} \\ \nonumber
K^{ij}  &=  \frac{1}{2}\left[\frac{f_{i}^{\prime\prime}(\bar x_1)}{f_i(\bar x_1)}\bar x_1^2+\frac{f_{j}^{\prime\prime}(\bar x_2)}{f_j(\bar x_2)}\bar x_2^2+2\bar x_1\bar x_2\frac{f_i^\prime(\bar x_1)f_j^\prime(\bar x_2)}{f_i(\bar x_1)f_j(\bar x_2)}\right]_{Q=M_V}
\end{align}
where  $f^\prime$ ($f^{\prime\prime}$) are the first (second) order derivatives\footnote{All the derivatives are evaluated as $f'=\frac{1}{\Delta x}\bigl[f(x+\Delta x/2)-f(x-\Delta x/2)\bigl]$, with a step of $\Delta x =10^{-5}$. This step has been varied in $\Delta x\in[10^{-6},10^{-4}]$ and the result is found to be stable in this range.} of the PDFs with respect to $x$. The constants $H_V$ and $K_V$ defined in Eq.~\ref{eq:second_order_formula} represent the appropriate average of the auxiliary functions of Eq.~\ref{eq:aux} over the flavour space.
The validity of the Taylor expansion of Eq.~\ref{eq:second_order_formula} has been assessed by comparing the lineshape from Eq.~\ref{eq:full_formula} and~\ref{eq:second_order_formula} at different values of $y$: the relative difference between the two is found to be below 0.5\% for $Q$ in a range of $\pm 2$ GeV around $M_V$.

In Eq.~\ref{eq:second_order_formula}, the contribution of the PDFs to the lineshape is fully encoded in the constants $H_V$ and $K_V$.
The mode $Q_0$ of the lineshape can be readily calculated from Eq.~\ref{eq:second_order_formula}:
\begin{equation}\label{eq:shift}
Q_{0} \approx M_V-\frac{\Gamma_V^2(H_V+1)M_V}{2\left[\Gamma_V^2(H_V+K_V)-4M_V^2 \right]} \approx M_V+\frac{\Gamma_V^2}{8M_V}(H_V+1),
\end{equation}
where the approximation $\Gamma_V^2(H_V+K_V)\ll 4M_V^2$ can be justified \textit{a posteriori}. The quantity
\begin{equation}\label{eq:shift2}
\Delta_V \equiv \frac{\Gamma_V^2}{8M_V}(H_V+1)
\end{equation}
represents the displacement of the mode $Q_0$ from $M_V$.
Part of it is simply due to the Jacobian factor from the transformation of Eq.~\ref{eq:change_var}, and does not depend on the PDFs.
The right-hand side of Eq.~\ref{eq:shift} depends on $K_V$ only at higher order in $\Gamma_V/M_V$ because it enters as the coefficient of a quadratic correction to the Breit-Wigner functions, which is symmetric around $M_V$.
Both $H_V$ and $K_V$ are functions of $y$ and $M_V$, albeit the dependence on the latter is negligible in the range of experimental uncertainty on $M_Z$ ($\sim 2$~MeV) and $M_W$ ($\sim 12$~MeV) compared to the PDF uncertainties on $\Delta_V$.

\begin{figure}[!htb]
\centering
\includegraphics[width=0.8\columnwidth]{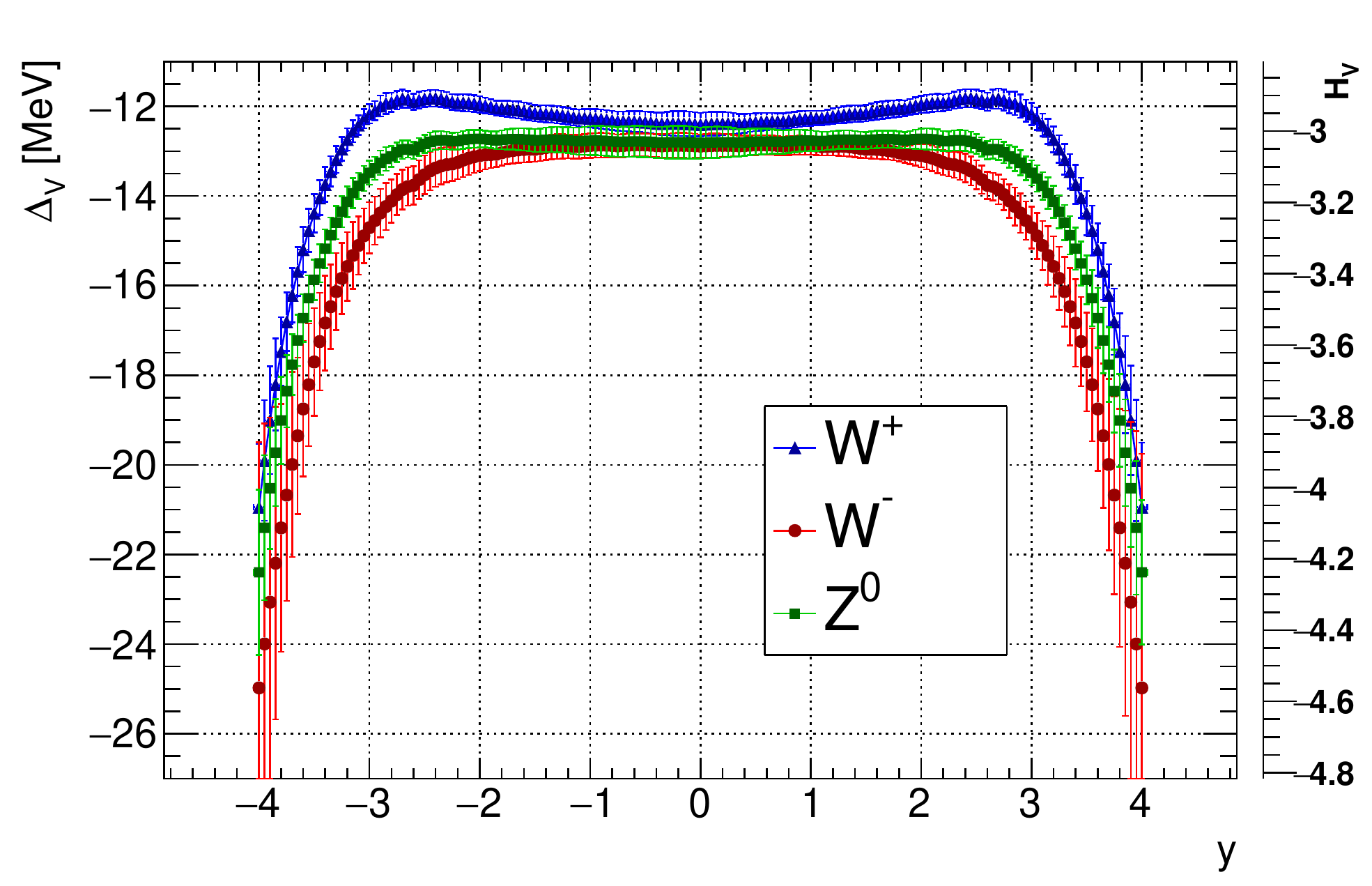}
\caption{The shift $\Delta_V$ and the coefficient $H_V$ as a function of $y$ averaged over the various quark flavours that enter the tree-level production of $W^\pm$ and $Z$ in $pp$ collisions at $\sqrt{s}=13$~TeV. On the right the equivalent scale for the correspondent $H_V$.}
\label{fig:shift_mean}
\end{figure}

The shift $\Delta_V$ determined from Eq.~\ref{eq:shift2} is plotted in Fig.~\ref{fig:shift_mean} as a function of the boson rapidity $y$ for $Z$ and $W^\pm$ production.
The error bars correspond to the RMS of the distribution obtained by sampling 100 replicas in the \texttt{NNPDF30\_nlo\_nf\_5\_pdfas} set from the LHAPDF library~\cite{LHAPDF}. A comparison with  \texttt{NNPDF31\_nlo\_pdfas}, which includes Drell-Yan measurements from the 8 TeV run of LHC \cite{Forte:NNPDF31}, has been performed as well, showing consistent results both in the central value and in the uncertainty.
Numerical values are reported in Table~\ref{tab:shift} for three representative values of $y$.
A negative shift with typical size $|\Delta_V|\sim 13$~MeV is observed in the central region $|y|\lesssim 3$, steeply increasing at larger rapidity values.
This behaviour can be understood qualitatively in terms of the valence quark density $xu_V$ and $xd_V$, which are typical benchmarks in PDF fits~\cite{NNPDF}.
Indeed, for any derivable and positive-definite function $f$, it holds that
\begin{equation}\label{eq:shift3}
\frac{xf^\prime}{f} = \frac{1}{f} \left( xf \right)^\prime - 1.
\end{equation}
The left-hand side of Eq.~\ref{eq:shift3} is of the same form of the terms that appear in the definition of $H_V$ (see the second line of Eq.~\ref{eq:aux}).
The valence quark densities feature a local maximum at $x\sim 10^{-1}$, which corresponds to $|y|\sim 3$ at $Q\sim 90$~GeV.
By identifying $f$ in Eq.~\ref{eq:shift3} with $u_V$ or $d_V$ one can easily see that the terms $\left( xf \right)^\prime/f$ vanish around $|y|\sim 3$,
thus giving the smallest shift, whereas they steeply decrease at higher rapidity values since $f\to 0$ and $(xf)^\prime$ becomes negative.
The relative PDF uncertainty on $\Delta_V$ is found to be in the 5\% ballpark, ranging from $0.3$ MeV at $|y|\sim0$ to $1$ MeV at $|y|\sim 3.5$.

A further inspection of Fig.~\ref{fig:shift_mean} shows that in the case of the $Z$ boson, the shift happens to lie between the shifts for the $W$ boson of opposite charge, as an effect of the different partonic favours probed by the gauge bosons.

Finally, the approximation which leads to Eq. \ref{eq:shift2} has been numerically validated. The resulting $K_V$ values range between 5 and 15, depending on $y$, with a relative uncertainty below 3\%.

\begin{table}[!htb]
\centering
\caption{Numerical values of the shift $\Delta_V$ in $pp$ collision at $\sqrt{s}=13$ TeV, for three selected values of $|y|$ with their PDF uncertainty.}\label{tab:shift}
\begin{tabular}{c  c c c  }
\hline\noalign{\smallskip}
$|y|$              &  $Z$ [MeV] & $W^{+}$ [MeV] & $W^{-}$ [MeV] \\
\hline\noalign{\smallskip}
0.0         & $-12.8 \pm 0.3$ & $-12.4 \pm 0.3$ & $-12.9 \pm 0.3$\\
2.0        & $-12.7 \pm 0.2$ & $-11.9 \pm 0.2$ & $-13.1 \pm 0.3$ \\
3.5		    & $-15.9 \pm 0.5$ & $-14.4 \pm 0.5$ & $-17.7 \pm 1.1$\\
\hline\noalign{\smallskip}
\end{tabular}
\end{table}

\section{Monte Carlo simulation study}\label{sec:MC}

The tree-level calculation of Sec.~\ref{sec:lhapdf} has been validated by using a MC simulation of $pp\to V+X$ production. Besides corroborating the tree level model, the MC analysis will also allow us to extend the study to the full phase-space, which includes the contribution of other diagrams.
Given the similarity between neutral- and charged-current Drell-Yan production, and the observation that $W$ boson production, split  by charge, can serve as a good proxy also for the $Z$ boson (see Fig.~\ref{fig:shift_mean}), and given the much larger MC sample available, the analysis has been restricted hereafter to the special case $V=W^\pm$ 
About $8\times10^7$ events in the final state ${W}^\pm \to \mu^\pm \nu_\mu$ have been generated using the \texttt{MG5\_aMC@NLO}~\cite{madgraph} program interfaced with \texttt{Pythia8}~\cite{Pythia}.
The dilepton mass is reconstructed using the muon momentum before QED final state radiation. The MC simulation is NLO accurate for observables inclusive in additional QCD radiation and it assumes $M_W^{\rm MC}=80.419$  GeV and $\Gamma_W^{\rm MC}=2.047$~GeV. As already anticipated in Sec.~\ref{sec:lhapdf}, the tree-level prediction is expected to be reproduced in the limit $q_{\rm T} \to 0$, where $q_{\rm T}$ is the transverse momentum of the  boson. Indeed, in this regime the relative contribution of the tree-level $2\to 1$ diagrams, which produce the boson at rest in the transverse plane, is enhanced compared to higher-order $2\to 2$ diagrams. In contrast, a reduction of the shift in the large $q_{\rm T}$ region is expected, where gluon-initiated diagrams dominate, thus reducing the sensitivity of the boson virtuality on the partonic luminosity.

\subsection{Fit to the MC sample}\label{sec:MC_fit}

The analytical study of Sec.~\ref{sec:lhapdf} shows that the shift in the MC sample has to be extracted from a statistical analysis of the dilepton mass distribution $d\sigma^{\rm MC}_W/dQ$. A crucial part of this task is to chose the correct functional form for $d\sigma^{\rm MC}_W/dQ$, capable of modelling the lineshape without introducing a bias in the estimator of $\Delta_V$. Motivated by the tree-level study, an \textit{ansatz} function of the same form of Eq.~\ref{eq:second_order_formula} has been chosen:
\begin{equation}\label{eq:formula3}
\frac{d\sigma^{\rm MC}_V}{dQ}(Q \, | \, y)=A\frac{Q^\alpha}{(Q^2-M^2)^2+M^2\Gamma^2} \left[1+H\left(\frac{Q}{M}-1\right)+K\left(\frac{Q}{M}-1\right)^2\right].
\end{equation}
The choice $\alpha=1$ defines the baseline function, which will be referred to as the \textit{modified Breit-Wigner}. Indeed, this functional form, which is identical to Eq.~\ref{eq:second_order_formula} from the tree-level study, will be explicitly validated by checking that the estimator of $M_W$ and $\Gamma_W$ is consistent with the input values of the MC simulation $M_W^{\rm MC}$ and $\Gamma_W^{\rm MC}$. As a further validation of this choice,  two alternative instances of the parametric family of functions in Eq.~\ref{eq:formula3} has been considered. The first is obtained by the choice $\alpha=H=K=0$, which reduces to a {Breit-Wigner}. This function is formally incorrect to model the dilepton mass distribution since it does not account for the Jacobian factor proportional to $Q$. However, it is a useful benchmark since it is symmetric around $Q=M_W$ so that the mass estimator also matches the mode.
The second alternative function is obtained by choosing $\alpha=1$ and $H=K=0$.
This function, which will be referred to as \textit{Breit-Wigner with Jacobian}, would be correct in the absence of the PDF distortion to the $Q$ distribution. It peaks at $Q\approx M_W+\frac{\Gamma^2}{8M_W}$, which is always larger than $M_W$.

Three statistical analyses of the simulated events have been performed. The first analysis is inclusive in the phase-space of the OLD{$W$} boson and allows us to benchmark the different fit functions with the largest possible statistical precision. The second analysis is differential in the boson rapidity and is expected to reproduce, at least qualitatively, the $y$-dependence from the tree-level model, as shown in Fig.~\ref{fig:shift_mean}. However, the comparison can only be approximate, since the latter predicts the transverse momentum $q_{\rm T}$ to be identically zero, whereas the MC simulation generates a physical spectrum of transverse momenta. The third analysis is performed in bins of $q_{\rm T}$, and is inclusive in $y$. It allows us to both validate the tree-level calculation of Sec.~\ref{sec:lhapdf} by extrapolating to $q_{\rm T}\to 0$, and to study the dilution effect at larger values of $q_{\rm T}$. The fit parameters of Eq.~\ref{eq:formula3} are determined by minimizing a $\chi^2$ constructed using the event counts in each bin of the histogram in the range $[79,82]$~GeV and the value of the fit function at the center of the bin.

\begin{figure}[!htb]
\centering
\includegraphics[width=0.85\columnwidth]{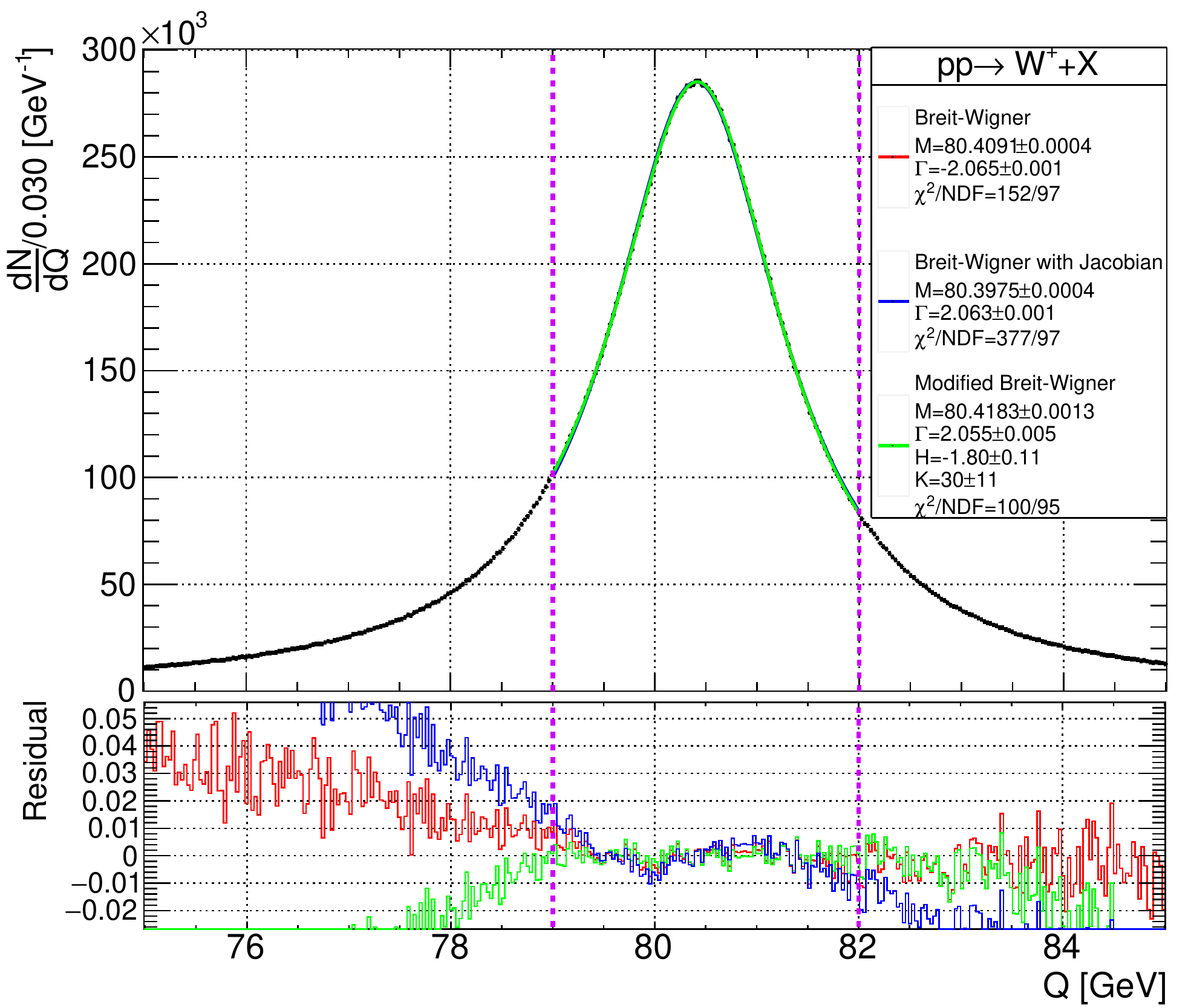}
\caption{The dilepton mass distribution for the inclusive $W^+$ sample. The result of the fit using the Breit-Wigner (red),  Breit-Wigner with Jacobian (blue), and {modified Breit Wigner} (green) are superimposed to the distributions. In the bottom pad, the residuals between the fitted function and the histogram are shown. Only events in the range $[79,82]$~GeV (marked by the vertical dashed lines) are used in the fit. A similar result is obtained for the $W^-$ sample.}
\label{fig:MC_integrated}
\end{figure}

Figure~\ref{fig:MC_integrated} shows the result of the three functional fits for the inclusive analysis for the $W^+$ sample. The $W^-$ sample shows very similar results for both the inclusive and the differential analyses, which are thus omitted for brevity. The quality of the fit improves dramatically when using the modified Breit-Wigner, with a reduced $\chi^2$ of about $1.0$ compared to $1.6$ and $3.9$ for the alternative functions. The best-fit value of $M_W$ when using the baseline function is consistent with the MC input within 1$\sigma$ ($1\pm 1$~MeV), whereas the alternative functions depart from it by $-9.9\pm 0.4$~MeV and $-21.5\pm 0.4$~MeV, respectively. Likewise, the best-fit value of $\Gamma_W$ is consistent with the MC input values within $1.5\sigma$ ($8\pm 5$~MeV) for the baseline function, while it departs from it by $18\pm1$~MeV and $16\pm 1$~MeV for the alternative functions. The residual discrepancy on $\Gamma_W$ when using the modified Breit-Wigner is ascribed to higher order terms in the power expansion, not included in Eq.~\ref{eq:formula3}. In order to validate this assumption, a toy MC has been used to verify that such a discrepancy is indeed consistent with the neglected terms that mostly enter through the tails of the distribution, whereas $M$, $H$ and $K$ are seen to be robust.

\begin{figure}[!htb]
\centering
\subfigure[\label{fig:MC_mass_y}] {\includegraphics[width=0.49\columnwidth]{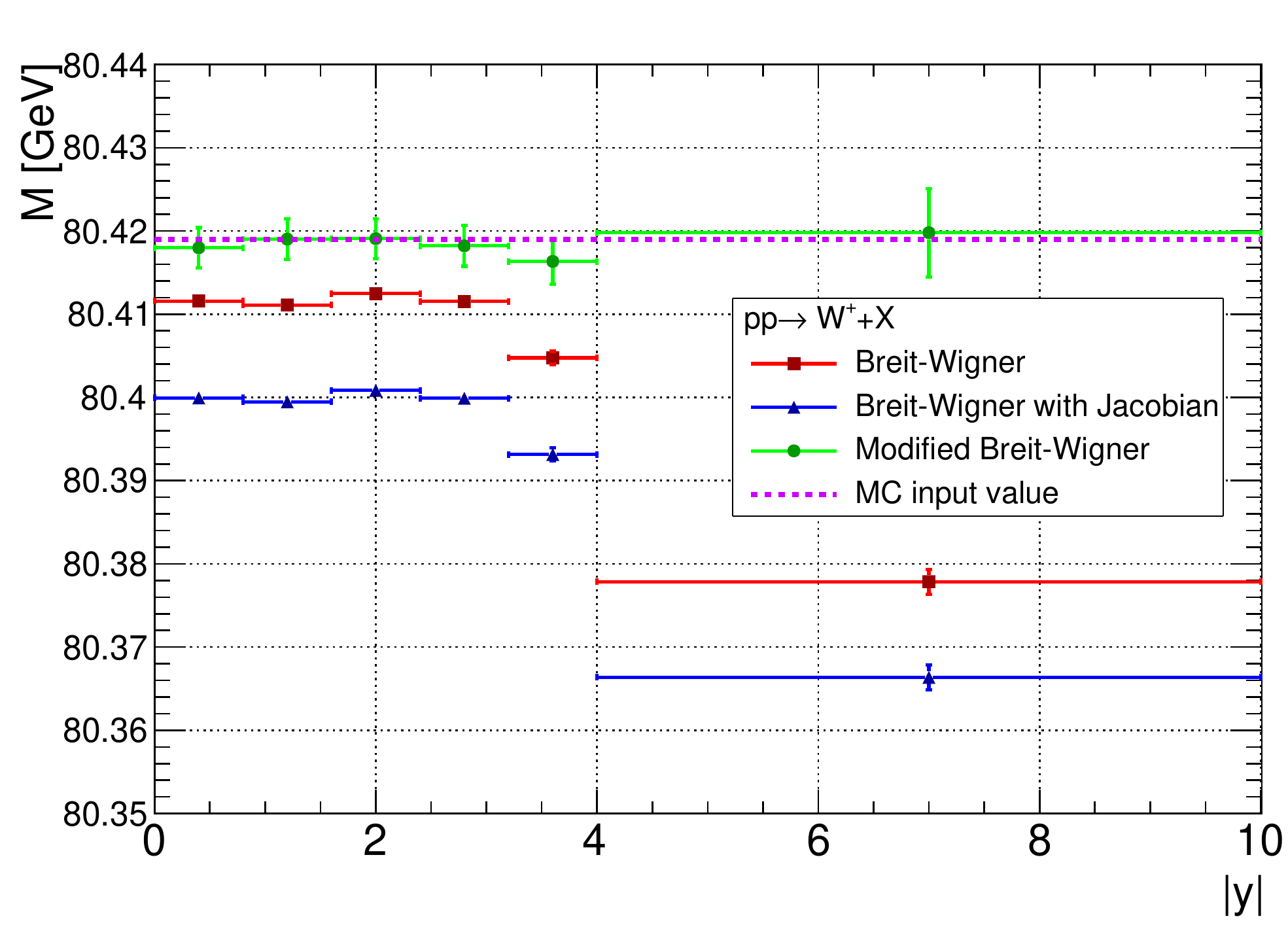}}
\subfigure[\label{fig:MC_mass_pt}]{\includegraphics[width=0.49\columnwidth]{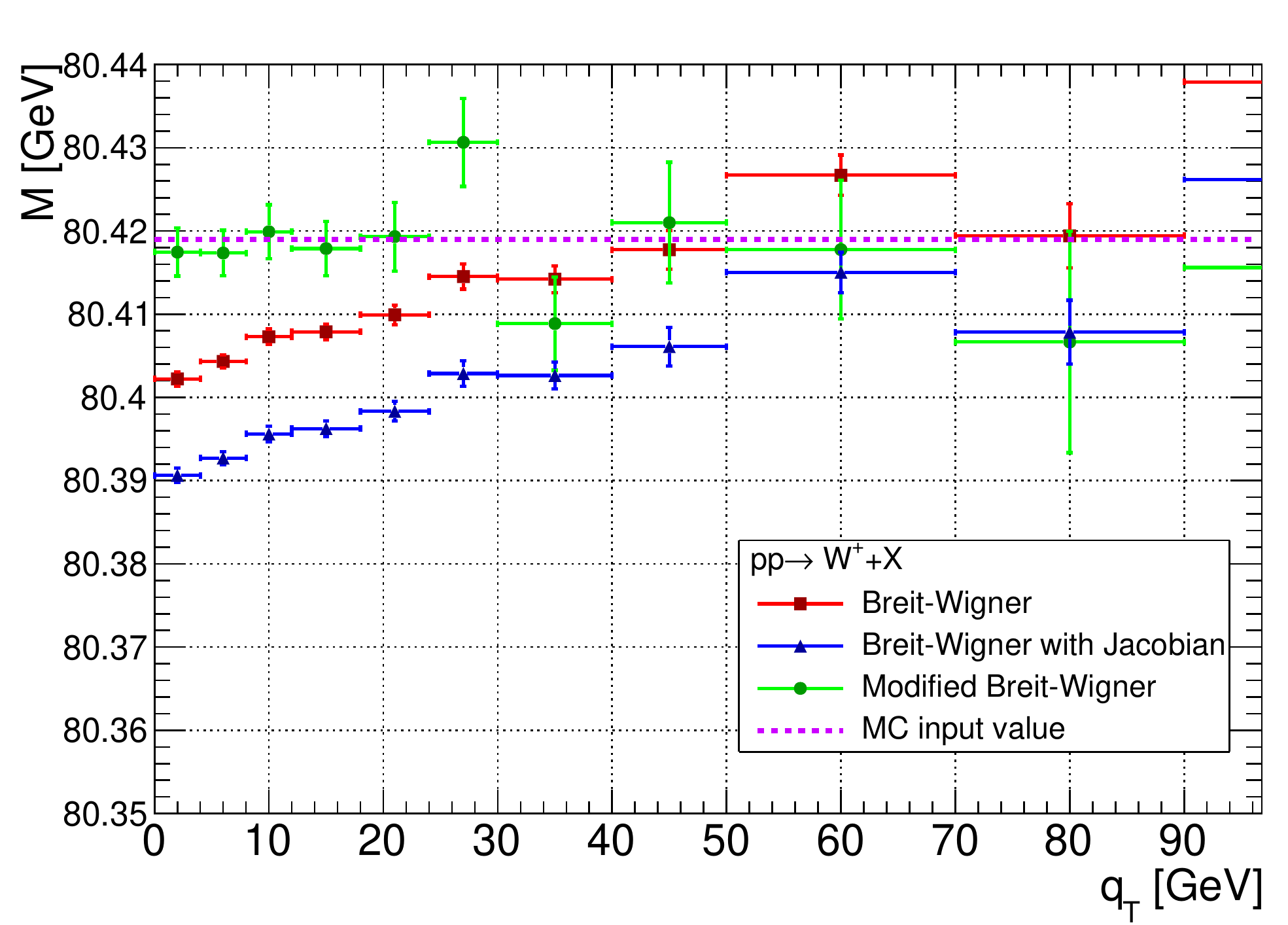}}
\caption{The best-fit value of $M_W$ using the Breit-Wigner (red),  Breit-Wigner with Jacobian (blue), and {modified Breit Wigner} (green), in bins of $|y|$ (left) and $q_{\rm T}$ (right), for the simulated $W^+$ sample. The dotted line corresponds to the input value of $M_W^{MC}$. A similar result is obtained for the $W^-$ sample. }\label{fig:MC_m}
\end{figure}

The best-fit value of $M_W$ from the differential analysis in the $W$ boson rapidity are reported in Fig.~\ref{fig:MC_mass_y} for the $W^+$ sample.
The {Breit-Wigner} fit underestimates $M_W$ all over the rapidity spectrum, as also observed in the inclusive analysis. The same applies to the {Breit-Wigner with Jacobian} function. For the latter, the discrepancy is even more pronounced. Indeed, the Jacobian factor contributes via a positive bias to the peak position. By neglecting the PDF term, which pulls in the opposite direction, the estimator of $M_W$ is thus shifted to even lower values compared to $M_W$. The {modified Breit-Wigner} function correctly reproduces the input value $M_W^{\rm MC}$ in all bins of $|y|$, including the high $|y|$ regimes, where the alternative functions perform rather poorly.

Finally, the results of the analysis differential in the $W$ boson transverse momentum are shown in Fig.~\ref{fig:MC_mass_pt}. The {modified Breit-Wigner} is seen to correctly reproduce the input mass value for all bins of $q_{\rm T}$, whereas the two alternative functions disagree, especially at low transverse momenta. For $q_{\rm T}$ in excess of about 40 GeV the statistical error is too large to discriminate among the models.

\subsection{Extraction of $\Delta_W$}\label{sec:MC_H}

Since the fit reproduces well the true values for $M_W$ and $\Gamma_W$, we fix the values of $M$ and $\Gamma$ to the MC input values in Eq.~\ref{eq:shift} and repeat the fit with $A, H, K$ as the only free parameters.

For the inclusive sample (Fig.~\ref{fig:MC_integrated}) the value of the shift is found to be:
\begin{equation}\label{eq:shift_integrated}
\text{(Full phase-space)}\qquad
\begin{aligned}
\Delta_{W^+}  & =-5.4\pm 0.2\,\text{(stat.) MeV} \pm 0.1 \, \text{(PDF) MeV},
\\
\Delta_{W^-} &=-5.8\pm 0.2\, \text{(stat.) MeV} \pm 0.1 \, \text{(PDF) MeV}.
\end{aligned}
\quad
\end{equation}
The first uncertainty is statistical-only while the second is the estimation of the systematic uncertainty from the PDFs. The latter is estimated from the RMS of the shifts determined using the first 100 replicas, as described in Sec.~\ref{sec:lhapdf}. In these fits the parameters $M$ and $\Gamma$ have been left free, since the uncertainty on the PDFs would be otherwise over-constrained by the imposed knowledge on the mass and the width of the resonance.

\begin{figure}[!htb]
\centering
\subfigure[\label{fig:MC_shift_y}]{\includegraphics[width=0.49\columnwidth]{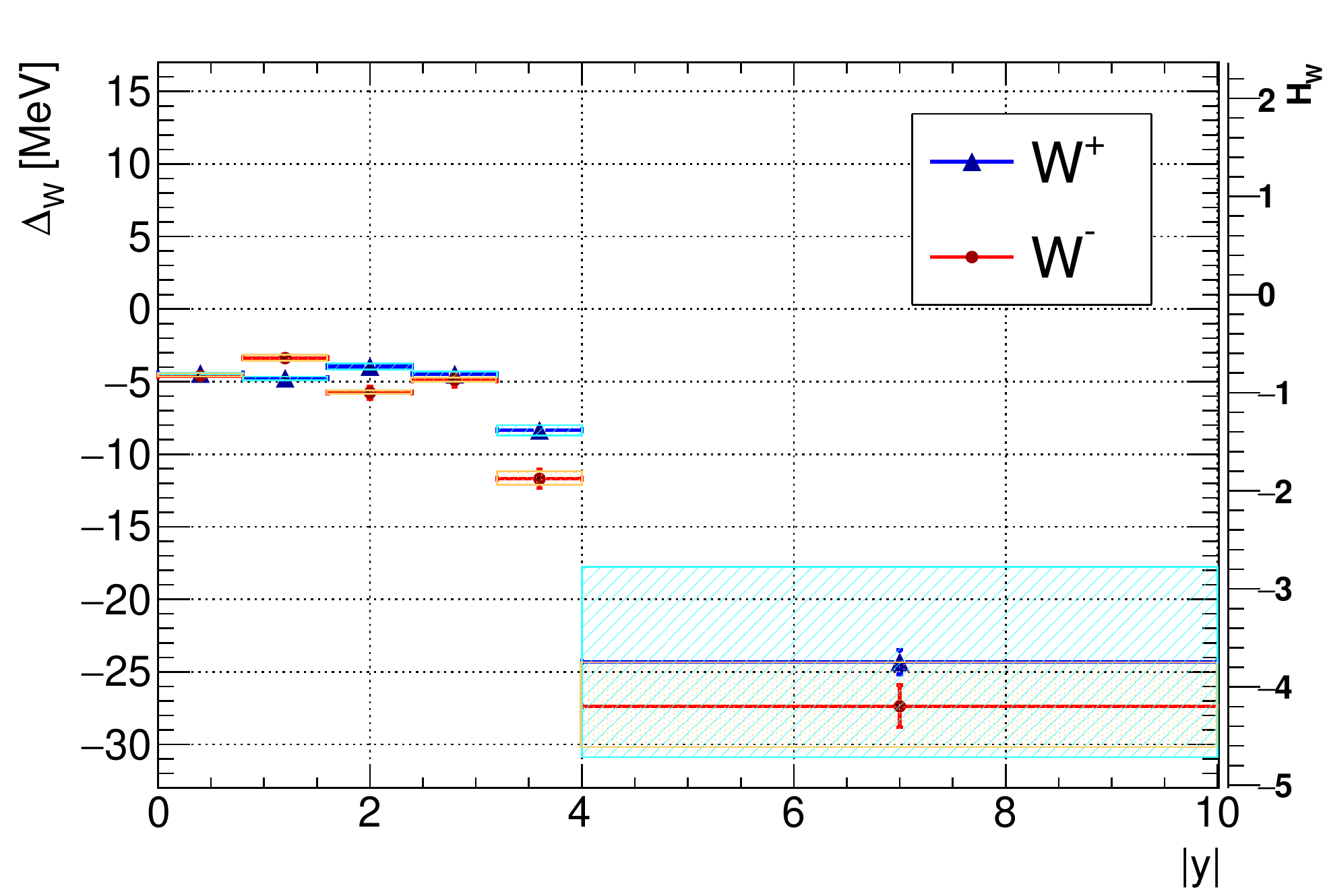}}
\subfigure[\label{fig:MC_shifty_pt}] {\includegraphics[width=0.49\columnwidth]{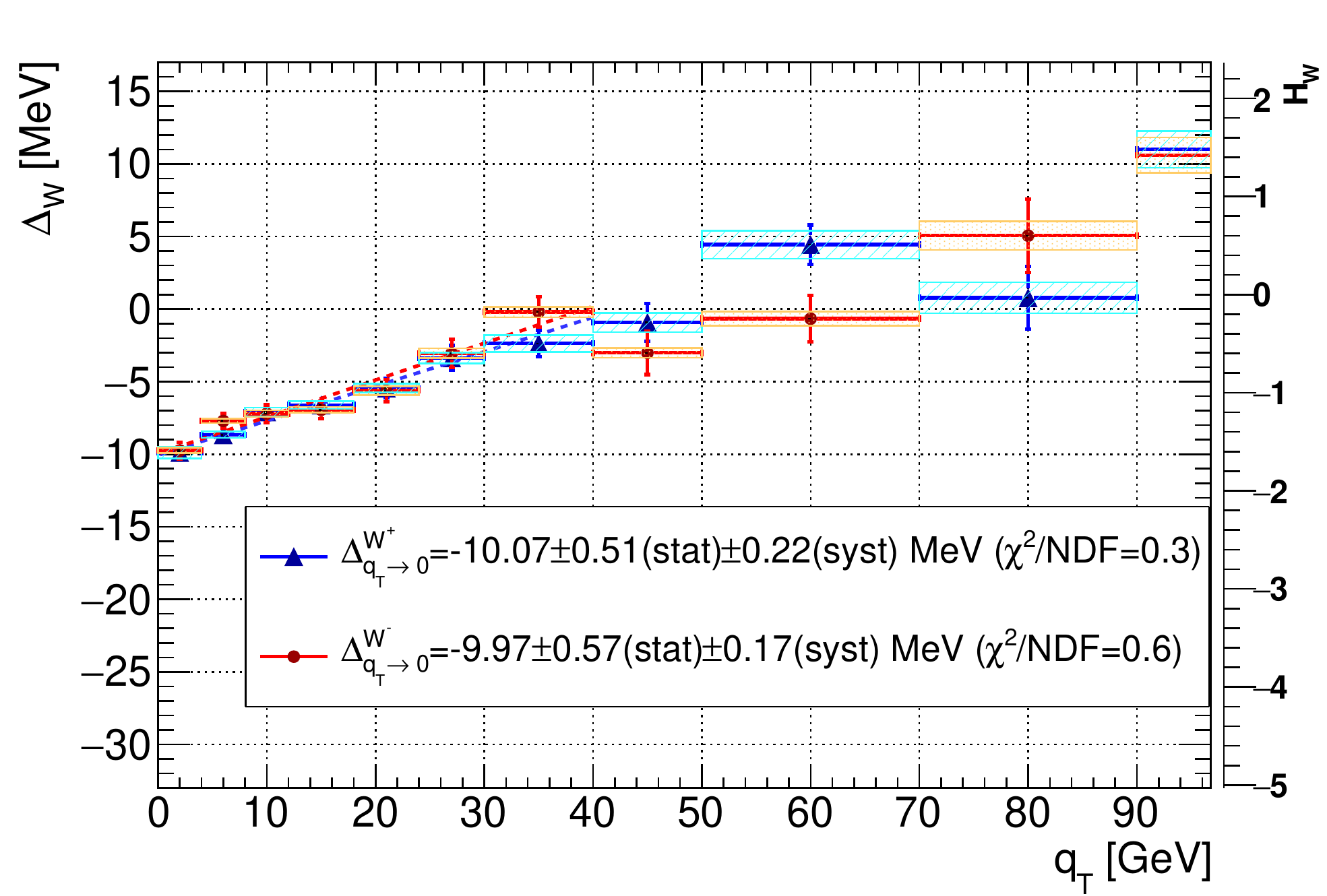}}
\caption{The shift $\Delta_{W^\pm}$ in bins of the $W$ boson rapidity $y$ (left) and transverse momentum $q_{\rm T}$ (right).
For the latter, a linear fit in the range $[0,40]$~MeV is performed to extrapolate the result to $q_{\rm T}\to 0$. The shaded boxes correspond to the PDF systematic uncertainty, as described in the text.
On the right side of each plot, the equivalent scale for the $H_W$ parameter is reported.}\label{fig:MC_H}
\end{figure}

The fitted values of $\Delta_W$ for the differential analyses are shown in Fig.~\ref{fig:MC_H} in bins of $|y|$ and $q_{\rm T}$, separately for $W^+$ and $W^-$.
The variation of $\Delta_W$ with the boson rapidity is shown in Fig.~\ref{fig:MC_shift_y}. It agrees well with the tree-level expectation of a flat shift in the central rapidity region followed by a rapid decrease at larger rapidity values. However, the shift in the central region is found to be smaller by a factor of about two, like for the inclusive results. Such a difference has been interpreted as the result of the dilution from higher-order diagrams. Indeed, in the limit $q_{\rm T}\to 0$, the measured shift gets closer to the tree-level result as shown by Fig.~\ref{fig:MC_shifty_pt}, while it vanishes for $q_{\rm T}$ in excess of about 40~GeV. A simple linear extrapolation to $q_{\rm T}\to0$ yields limiting values of
\begin{equation}\label{eq:shift_extrap}
\text{(}q_T\to 0\text{ extrapolation)}\qquad
\begin{aligned}
\Delta_{W^+} & =-10.1 \pm 0.5 \, \text{(stat.)} \pm 0.2 \, \text{(PDF) MeV}, \\
\Delta_{W^-} & =-10.0 \pm 0.6\,\text{(stat.)} \pm 0.2 \, \text{(PDF) MeV}.
\end{aligned}
\end{equation}

Although reasonably close to the tree-level expectation, this result still disagrees with it by roughly 30\%.
This residual difference is interpreted as a pure next-to-leading-order correction to the leading-order prediction, stemming from collinear gluon emission and from gluon-initiated diagrams which contribute to the small-$q_{\rm T}$ regime. The relative PDF uncertainty is found to agree reasonably well with the expectation from the tree-level model averaged over the $W$ boson rapidity.

As a cross-check of this result, the fit has been repeated varying the range symmetrically by $\pm10\%$. The results for $\Delta_W$ are stable, with a maximum discrepancy of 5\%, which is within the uncertainty of the parameter. The fit has been also repeated after changing the renormalization and factorization scales in the matrix elements of the MC simulation by factors of $0.5$ and $2$, respectively. The results are again found to be stable within the PDF uncertainty. 

Due to the agreement between the MC studies above and tree-level model of Sec.~\ref{sec:lhapdf},
 the results for $\Delta_W$ are expected to be valid also for the $Z$ boson case with $\Delta_{W^-}<\Delta_Z<\Delta_{W^+}$ and a similar PDF-related uncertainty.

\section{Conclusions}\label{sec:conclusion}

In this paper, the impact of the PDFs on the lineshape of  gauge bosons at the LHC has been investigated. Given the narrow width of the electroweak gauge bosons, the PDF impact can be treated, to a first approximation, as a shift $\Delta_V$ of the mode of the dilepton mass spectrum from the boson mass $M_V$. The origin of such shift can be traced back to the dependence of the partonic luminosity on the virtuality $Q$ of the gauge boson. 
This effect is particular important for a possible future precise measurement of the $Z$ boson mass since it  directly affects the extraction of $M_V$ from the kinematics of the dilepton final state.
It has been first studied analytically by using a tree-level model of Drell-Yan production and then validated by a statistical analysis of a MC simulated sample. The tree-level calculation agrees reasonably well with the MC study in the phase-space where the two are expected to be comparable. The results of this study prove that the PDF uncertainty on $\Delta_V$ is below one MeV all over the phase space relevant for future mass measurements at the LHC.

\begin{acknowledgements}
This work has been partially supported by MIUR, PRIN 2017, through the PRIN 2017F28R78 Project. In addition we wish to acknowledge the help of Paolo Azzuri, from INFN - Sezione di Pisa, in reading of the final draft.
\end{acknowledgements}


\bibliographystyle{spphys}       
\bibliography{bibliografia.bib}   

\end{document}